\documentclass[aps,pra,amsmath,amssymb,showpacs,twocolumn]{revtex4}

\usepackage{amssymb}
\usepackage{amsmath}
\usepackage{amscd}
\usepackage{natbib}
\usepackage{graphicx}            

\newcommand{\be}{\begin{equation}}
\newcommand{\ee}{\end{equation}}

\newcommand{\ab}{\hspace{-1.5mm}}

\newcommand \pp{{\bf p}}
\newcommand \qq{{\bf q}}

\newcommand \vv{{\bf v}}

\renewcommand{\vec}[1]{{\mathbf #1}}

\begin{document}


\title{Polar molecules in bilayers with high population imbalance}

 \author{Michael Klawunn and Alessio Recati}
 \affiliation{
 \mbox{INO-CNR BEC Center and Dipartimento di Fisica, Universit\`a di Trento, 38123 Povo, Italy}}

\date{\today}

\begin{abstract}
We investigate a dilute Fermi gas of polar molecules confined into a bilayer setup with dipole moments polarized perpendicular to the layers.
In particular, we consider the extreme case of population imbalance,
where we have only one particle in one layer and many particles in the other one.
The single molecule is attracted by the dilute Fermi-gas through the inter-layer dipole-dipole force presenting
an interesting impurity problem with longrange anistropic interaction.
We calulate the chemical potential of the impurity, in second order perturbation theory and in ladder approximation with a Brueckner-Hartree-Fock approach. Moreover, we determine the momentum relaxation rate of the impurity, which is related to the ``dipolar" drag effect.
For a confined system we relate the results for the chemical potential with the measurement of the
collective modes of the impurity. The momentum relaxation rate provide instead an estimate on how quickly the oscillations are damped.
\end{abstract}


\maketitle

\section{Introduction}

During the last years ultracold dipolar gases have attracted great interest because the dipole-dipole interaction (DDI)
drastically changes the nature of quantum degenerate regimes compared to ordinary short-range interacting
gases \cite{Baranov2008,Lahaye2009}. 
In recent experiments polar molecules in the ground ro-vibrational state has been created \cite{Ni2008, Deiglmayr2008} and
cooled towards quantum degeneracy \cite{Ni2008}.
The main obstacle is the decay of the system due to ultracold chemical reactions found in recent experiments \cite{Ospelkaus2010}.
However, if the molecules are confined in a 2D geometry and oriented perpendicularly to the plane of their motion by a strong external electric field,
these reactions are expected to be suppressed by the intermolecular repulsion \cite{Ni2010}.
Bilayer arrangements with this dipolar orientation are particularly interesting,
since they allow for both, stability against chemical reactions and effects induced by the anisotropy of the DDI.
The attractive inter-layer interaction gives rise to a peculiar two dimensional
scattering behaviour and bound states \cite{Michael2D,Armstrong2010,Baranov2011}
and may induce interesting phenomena like inter-layer pairing and superfluidity \cite{BCS-bilayer,Baranov2011,Potter2010}
and non-local state-changing collisions \cite{Pikovski2011}.
Notice that the bilayer system can be also thought as a single layer system with two different species interacting through the intra-layer and the inter-layer DDI. Until now only the balanced situation with an equal number of particles in both layers has been studied.

\begin{figure}[t]
\begin{center}
\includegraphics[width=0.44\textwidth,angle=0]{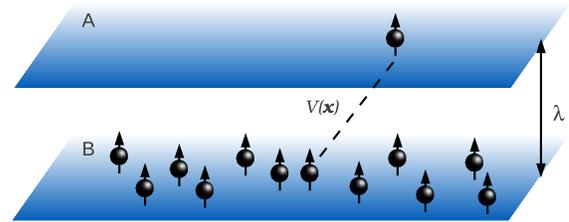}
\caption{Scheme of the system under consideration.}
\label{fig:scheme}
\end{center}
\end{figure}

For atomic short-range interacting Fermi gases of different species (e.g. atoms in two different hyperfine states) however,
imbalanced and in particular highly imbalanced gases have been extensively studied experimentally \cite{ExpPol} and theoretically (see, e.g., the recent review \cite{ChevyMora} and references therein). The building block for the understanding of such systems 
is the solution of the limiting case of a single impurity atom interacting via a short range potential with an ideal atomic Fermi gas.
Such a problem is not only relevant in the field of ultra-cold gases,
since it is related to the so-called impurity problem, which is present also in other areas of physics.
Analogously it is interesting to study the bilayer system of polar molecules with a high population imbalance, resembling an impurity
interacting via the inter-layer DDI with a Fermi gas (see Fig. \ref{fig:scheme}). In the following we call the single particle in one layer impurity 
or Fermi-polaron or polaron in analogy with electrons in a crystal dressed by the bosonic (phonon) bath.

An important quantity characterizing the impurity is its chemical potential, namely 
the (in our case negative) energy difference of the ground state with and without
the impurity at rest. We will sometimes call this quantity also interaction energy or binding energy.
For short-range interacting Fermi gases this quantity has been successfully calculated with various methods
in one \cite{McGuire,Roland1D}, two \cite{var2D-1,var2D-2,OurPRA,Schmidt2D} and three dimensions \cite{Alessio2007,PilatiGiorgini,Schmidt2011}.

In this paper we investigate the problem of an impurity in one layer ($A$) interacting via the DDI with the Fermi gas phase in the other layer ($B$) as sketched in Fig. \ref{fig:scheme}).
Assuming no inter-layer tunnelling we calculate the chemical potential of the impurity in second order perturbation theory and in ladder approximation
using a Brueckner Hartree Fock approach.
It has been shown in \cite{Alessio2007,Lipparini} that for short range interacting atomic gases the latter is equivalent
to a self-consistent T-matrix approach and a variational ansatz, which have been applied succesfully to the polaron problem.
Moreover, we study some dynamical properties of the impurity: we estimate the momentum relaxation rate  and calculate the eigenfrequency of the impurity in an external trap. 
Note that we consider through the whole paper weak interactions and a dilute Fermi gas of polar molecules. 

The structure of the paper is as follows: in section \ref{sec:IIorder} we introduce the relevant physical quantities and study the polaron energy in second order
perturbation theory. In section \ref{sec:BHF} we extend our analysis to higher orders and calculate the chemical potential in ladder approximation within
Brueckner-Hartree-Fock approach. In section III we estimate the momentum relaxation time of a moving impurity. Finally in section IV we use our results for homogeneous systems to investigate the motion of the impurity in an external trap using the local density approximation.

\section{Chemical potential}

\subsection{Second order perturbation theory}\label{sec:IIorder}

The model Hamiltonian can be written as
\begin{align}
\begin{split}
H=&\sum_{\vec{k},\sigma} \epsilon_k c^\dagger_{\vec{k},\sigma} c_{\vec{k},\sigma}\\
 &+\frac{1}{2V}\hspace{-0.5cm}\sum_{\vec{p_1},\vec{p_2},\vec{q},\sigma,\sigma'}\hspace{-0.5cm} V^{\sigma\sigma^\prime}(\vec{q})
c^\dagger_{\vec{p_1}+\vec{q},\sigma} c^\dagger_{\vec{p_2}-\vec{q},\sigma'} c_{\vec{p_2},\sigma'} c_{\vec{p_1},\sigma},
\end{split}
\end{align}
where $\epsilon_k$ is the single particle energy and $c^\dagger_{\vec{p}\sigma}$ ($c_{\vec{p}\sigma}$)
creates (annihilates) a fermionic particle with momentum $\vec{p}$ in layer $\sigma=\left\lbrace A,B\right\rbrace$. 
The inter-layer and intra-layer dipolar interaction potentials are $V^{\sigma\sigma}$ and $V^{AB}$,
 respectively.
Note that up to second order the inter-layer and the intra-layer interaction decouple,
such that the energy of the impurity does not depend on the intra-layer interaction. At higher orders, processes mediated by the intra-layer interaction
might become important and have to be taken into account in the calculation of the chemical potential of the impurity.
Thus in the following second order calculation we consider only the inter-layer interaction and we can omit the indices of the potential. 
The intra-layer interaction energy has been calculated in first order perturbation theory in Refs. \cite{Baranov2011,inlayer}

For two molecules with mass $m$ and dipole moment $d$ placed in two layers at distance $\lambda$ 
the DDI potential reads
\begin{align}\label{VAB}
V^{AB}(r) =d^2 \frac{r^2-\lambda^2}{\left( r^2+\lambda^2\right) ^{5/2}},
\end{align}
with $r$ being the relative distance in the plane of motion, whose Fourier transform is
\begin{align}\label{V}
V(\vec{q}) \equiv V^{AB}(\vec{q}) = -2\pi d^2 q e^{-\lambda q}.
\end{align}

In first order perturbation theory the interaction energy reads $E^{(1)}=\left\langle GS | H^{AB} |GS \right\rangle $,
where $|GS \rangle = | \left\lbrace n_{\vec{p_1},B} \right\rbrace,  \vec{P}  \rangle$ is the ground state
of the system with $N_B$ particles in layer $B$ and one particle (the impurity) in layer $A$ with momentum $\vec{P}$.
The first order contribution \be E^{(1)}=n_B V(q=0)=0 \ee where $n_B=N_B/V$ is the particle density in layer $B$, vanishes. Indeed the potential is partitially attractive and partially repulsive such that it satisfies $\int d \vec{r} V^{AB}(\vec{r})=0$.
Similiarly, the second order contribution can be found as
\begin{align}\label{E2}
E^{(2)}= \sum_\vec{p_1} n_{\vec{p_1},B} \frac{m}{V}\sum_{\vec{q}\neq 0}
\frac{2|V(\vec{q})|^2 (1 - n_{\vec{p_1}+\vec{q},B})}{\vec{p_1}^2+\vec{P}^2-(\vec{p_1}+\vec{q})^2 - (\vec{P}-\vec{q})^2 },
\end{align}
where $n_{\vec{k},B}=\Theta(k_{F,B} - k)$ denotes the Fermi-distribution at zero temperature. At rest ($P=0$) this energy is
the chemical potential of the impurity and in Sect. \ref{sec:BHF},
we discuss it as a function of the dipole moment $d$, the inter-layer distance $\lambda$
and the Fermi-momentum $k_{F,B}$ of the gas in layer $B$.

\subsection{Brueckner-Hartree-Fock approach}\label{sec:BHF}

We calculate the polaron energy summing up the ladder diagrams for the inter-layer interaction only,
which at low enough density (in layer $B$) should give the most relevant contribution.
The basic equation is the Bethe-Goldstone integral equation
for the reaction matrix \cite{BG1957}, also called effective interaction
(e.g., for the 2D electron gas \cite{Nagano1984}). 
In our case the Bethe-Goldstone integral equation for the effective interaction between a particle in layer $B$ with momentum $\vec{k_1}$ and 
the impurity atom in layer $A$ with momentum $\vec{k_2}$ can be written as
\begin{align}\label{BG}
\begin{split}
&g(\epsilon_p, \vec{k_1},\vec{k_2};\vec{q})=V(\vec{q})+\int \frac{d\vec{k}}{(2\pi)^2}
V(|\vec{q}-\vec{k}|)\times\\
& \frac{2m\left(1-n_{\vec{k_1}+\vec{k},B}\right) }
{2m\epsilon_p + k_1^2+k_2^2-(\vec{k_1}+\vec{k})^2
-(\vec{k_2}-\vec{k})^2}
g(\epsilon_p, \vec{k_1},\vec{k_2};\vec{k})
\end{split}
\end{align}
In Eq. (\ref{BG}) $\vec{q}$ is the momentum transfer, $V(\vec{q})$ is the Fourier transform
of the inter-layer potential (\ref{V}), $n_{\vec{k}}$ the Fermi distribution function at
zero temperature.
Note, that the interaction energy $\epsilon_p$ has been included in the initial energy of excitation processes
$\epsilon_p + \frac{k_1^2}{2m}+\frac{k_2^2}{2m}$.
This interaction energy or correlation energy follows then selfconsistently from the mean value of the effective interaction
$\epsilon_p= \left\langle g(\epsilon_p,\vec{k_1},\vec{k_2};\vec{q})\right\rangle$

For  $\vec{k_2}=0$ one gets the rest correlation energy $\epsilon_p^0$
of the polaron, and by expanding this solution in $k_2^2$ one get its effective mass $m^*$ as usual
by the relation $E = \epsilon_p^0+k_2^2/2m^*$.  
We remind that in Eq. (\ref{BG}) for the effective interaction only ladder diagrams are 
summed and the Fermi sea limits the momenta in the intermediate states.

For shortrange interacting Fermi gases the integral equation (\ref{BG}) cannot be solved directly
due to the infinite hard core of the potential. Especially in 2D one has to renormalize the equation by replacing
the potential through the off-shell scattering amplitude and solve the renormalized equation. We followed this strategy in
Ref. \cite{OurPRA}. However, the function of the inter-layer DDI $V(q)$ as given by Eq.(\ref{V}) goes to zero
for small as well as for large arguments, such that it does not cause any divergencies.
This allows for a direct numerical solution of Eq. (\ref{BG}) in our case.
Then we can compare this solution with the one obtained by means of the renormalization strategy as discussed in \cite{OurPRA}.

For our problem the Bethe-Goldstone equation (\ref{BG}) is a two-dimensional integral equation in the last argument
of $g(\epsilon_p, \vec{k_1}, \vec{k_2};\vec{q})$
which for $k_2=0$ can be written in units of $k_{F,B}$ as a Fredholm integral equation of the second kind
\begin{align}\label{BGnum}
\begin{split}
&g(\epsilon_p^0, k_1; q,\theta)=\tilde{V}(q)+\\
&\int_0^{\infty} \frac{dk}{(2\pi)} \int_0^{2\pi} \frac{d\phi}{(2\pi)}
K(\epsilon_p^0, k_1; q,\theta, k,\phi) g(\epsilon_p^0, k_1; k,\phi),
\end{split}
\end{align}
with the angles $\theta=\measuredangle(\vec{k_1},\vec{q})$ and $\phi=\measuredangle(\vec{k_1},\vec{k})$. In Eq. (\ref{BGnum}) we have introduced 
the dimensionless inter-layer dipolar potential 
\begin{align}\label{V1}
\tilde{V}(\vec{q}) = -2\pi (r^* k_{F,B}) e^{-(\lambda k_{F,B}) q},
\end{align}
with $r^*=d^2 m/\hbar^2$ the length scale associated with the dipolar interaction and 
the kernel
\begin{align}\label{K}
K(\epsilon_p, k_1; q,\theta, k,\phi)=
\frac{k \tilde{V}(|\vec{q}-\vec{k}|) \left\{1-\Theta(1-|\vec{k_1}+\vec{k}|) \right\}}{\epsilon_p^0/(2\epsilon_{F,B}) -k^2-k k_1 \cos\phi }.
\end{align}

The mean value of the solution of Eq.(\ref{BGnum}) gives the chemical potential $\tilde{\epsilon}_p^0=\epsilon_p^0/(2\epsilon_{F,B})$ of the impurity self-consistently
\begin{align}\label{E}
\begin{split}
\tilde{\epsilon}_p^0=&\int_0^{1} \frac{dk_1 k_1}{(2\pi)} \ab\: \int_0^{\infty} \ab \frac{dq q}{(2\pi)}
\ab\: \int_0^{2\pi} \ab \frac{d\theta}{(2\pi)} \: \tilde{V}(q) \times \\
 &\frac{1-\Theta(1-|\vec{k_1}+\vec{q}|)}{\tilde{\epsilon}_p^0 -q^2-q k_1 \cos\phi }\:\:
g(\tilde{\epsilon}_p^0, k_1; q,\theta).
\end{split}
\end{align}

We can take as quantities characterizing the inter-layer DDI the dipolar strength $r^*k_{F,B}$
and the inter-layer distance $\lambda k_{F,B}$.
The quantity $r^*k_{F,B}$ is also the strength of the intra-layer interaction in layer $B$ for which we consider only small values, such that we are far away from the liquid-crystal phase transition \cite{Astra2007}. In this case the intra-layer interaction could be taken into account by introducing renormalized Fermi liquid Landau parameters, which accounts, e.g., for an effective mass of the fermions in layer $B$ \cite{Baranov2011}.

Fig. 2 shows the numerical solution in ladder approximation (Eqs. (\ref{BGnum}),(\ref{K}) and (\ref{E})) for $r^*k_{F,B}=0.5$.
The polaron energy is plotted as a function of the inverse inter-layer distance $1/(\lambda k_{F,B})$. For comparison the second order perturbation theory result as obtained from Eq. (\ref{E2}) 
is also reported.
\begin{figure}[ht]
\begin{center}
\includegraphics[width=0.44\textwidth,angle=0]{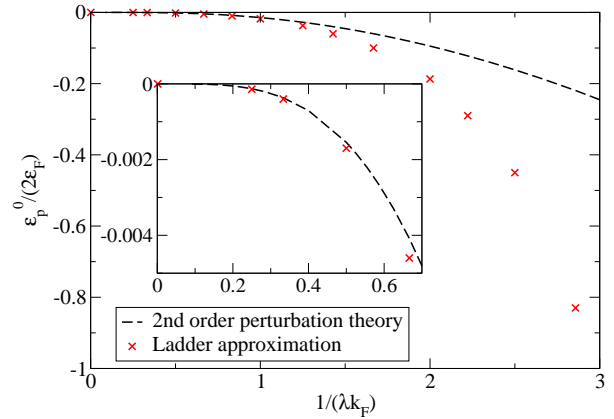}
\caption{Main: polaron energy as a function of the inverse inter-layer distance for dipolar strength of
$r^*k_{F,B}=0.5$ calculated in second order perturbation theory (black dashed line) and in ladder approximation (red crosses).
Inset: Zoom onto the weakly interacting regime, where second order perturbation theory is a good approximation.}
\label{fig:Ep1}
\end{center}
\end{figure}

As mentioned above one can also solve Eq. (\ref{BG}) approximately by replacing the potential through the scattering amplitude.
In Ref. \cite{OurPRA} we obtained in that way for the shortrange interacting Fermi gas an approximate analytical solution for the polaron energy.
The energy does not depend on the particular form of the interaction potential but only on the binding energy of the two-body bound state.
We can apply the results of \cite{OurPRA} for all potentials providing a normal scattering behaviour,
governed in 2D by a logarithmic on-shell scattering amplitude. Scattering and bound states of the inter-layer dipolar potential has been analysed
with different methods in detail in \cite{Michael2D,Baranov2011} and one has
that for a certain regime of parameters $r^*$ and $\lambda$ the low-energy scattering is indeed of the normal logarithmic form. In this case the polaron energy reads \cite{OurPRA}
\begin{align}\label{E_int_bound}
\epsilon_p^0 =\frac{-2\epsilon_{F,B}}{\ln\left[1+\frac{2\epsilon_{F,B}}{|\epsilon_b|} \right]},
\end{align}
where $|\epsilon_b|$ is now the binding energy of the two-body bound state of the inter-layer potential Eq. (\ref{V}).
This binding energy depends on $r^*$ and $\lambda$ and has been calculated in \cite{Michael2D} and \cite{Baranov2011} as well.
Using this results we can calculate the polaron energy from Eq. (\ref{E_int_bound}) in the regime where the inter-layer potential 
provides a logarithmic low energy scattering amplitude. This condition is satisfied for $r^*/\lambda \gtrsim 0.7$  \cite{Michael2D}.
Fig. 3 illustrates the comparison of the results obtained from the numerical solution of Eqs. (\ref{BGnum}),(\ref{K}) and (\ref{E})
and formula (\ref{E_int_bound}) for $r^*k_{F,B}=0.5$ 
and shows moreover the two-body binding energy for the inter-layer potential as obtained in \cite{Michael2D}.

Since the self-consistent approach we use in this paper is equivalent to a variational ansatz with a polaron many body wave function,
we expect that it cannot produce accurate results in the regime of strong two-particle interaction, where the particles are deeply bound in a
molecular state. The investigation of this molecular regime is beyond the scope of our paper. Moreover it has been shown for short-range interacting Fermi gases that the molecular regime in 2D, differently to the three-dimensional case, occurs for very large interaction strength \cite{var2D-2,Giorgini2D}. For our case strong particle interaction means that for fixed $r^*k_{F,B}$ the inverse inter-layer distance 
is very large. We find that for $1/(\lambda k_{F,B}) \gg 1$ the polaron energy converges to $\epsilon_p^0 \approx |\epsilon_b| - 0.55\epsilon_{F,B}$,
which is in good agreement with the result in \cite{var2D-2} for the polaron ansatz in the two-dimensional shortrange interacting gas.

\begin{figure}[]
\begin{center}
\includegraphics[width=0.44\textwidth,angle=0]{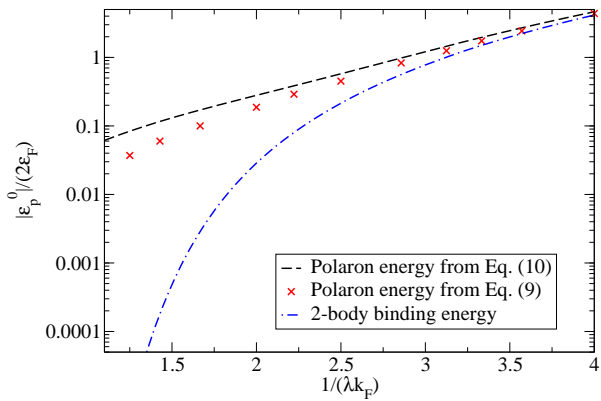}
\caption{Polaron energy in log scale as a function of the inverse inter-layer distance for a dipolar strength of
$r^*k_{F,B}=0.5$ calculated from Eq. \ref{E} (red crosses) and from Eq. \ref{E_int_bound} (black dashed line). 
The blue dashed dotted line indicates the two-body binding energy $\epsilon_b$ for the inter-layer DDI potential (\ref{V})
as calculated in \cite{Michael2D}}.
\label{fig:Ep2}
\end{center}
\end{figure}

\section{Momentum relaxation rate}

In the previous section we have calculated the energy of the polaron as a function of the interaction strength. We assumed the polaron is well defined with an energy given at finite momentum $\epsilon_q=\epsilon_p^0+q^2/(2 m^*)$, where $m^*\simeq m$ in the regime we have studied. 

Here we concentrate on estimating the finite momentum relaxation rate, $1/\tau_P$, of a polaron (gas), due the dipole interaction between the two layers, as a function of its momentum and of the temperature of the system. 
Such a quantity plays an important role in determining the transport properties -- in particular the cross resistance -- in double-layer semiconductor structures (see e.g., \cite{Rojo} and reference therein) or double-layer graphene systems (e.g., \cite{SarmaGraphene}). 
Indeed in coupled two-dimensional electron gases the electron-electron scattering between the two layers gives rise to the Coulomb drag effect, where a ``drag" current
is induced purely from the momentum exchanges through inter-layer electron-electron scattering events.
In a different context the polaron momentum relaxation rate (or spin-drag \cite{Polini}) has already been calculated for a three-dimensional highly polarized spin-$1/2$ Fermi gases, where the interaction is short range \cite{GeorgAlessio}. 

In a real polaron experiment one will have in the upper layer a gas at very low density $n_A$, much smaller than the lower layer density $n_B$. We take the gas in $A$ to have a momentum per unit volume ${\mathbf P}_A=n_A m {\mathbf v}$, with ${\mathbf v}$ the average velocity of the gas. We also assume that in both layers the gases are described by free quasi-particles in thermal equilibrium at temperature $T$ whose distribution functions are
$n_{\pp B}=f[\beta(\epsilon_{\pp B}-\mu_{B})]$
and $n_{{\mathbf{p}}A}=f[\beta(\epsilon_{\pp A}-{\mathbf
{p}}\cdot{\mathbf {v}}-\mu_A)]$, with $\beta=1/k_BT$ and
$f(x)=1/({\rm e}^x+1)$ the Fermi distribution. The single particle energies are
$\epsilon_{\pp A (B)}=p^2/2m$. The term ${\mathbf {p}}\cdot{\mathbf {v}}$ boosts
the $A$-layer distribution function by a velocity ${\mathbf{v}}$.

The momentum relaxation rate $\tau_P$ is defined by 
\be
\frac{{\mathbf P}_A}{dt}=-\frac{{\mathbf P}_A}{\tau_P}.
\ee
We will explicitly give the expressions of $\tau_P$ for relevant cases, closely following the calculation for the Coulomb drag between parallel two-dimensional electron gas by Jauho and Smith \cite{JauhoSmith}.

The rate of change of the momentum of the dipolar gas in layer $A$ due to the scattering from atoms/molecules in the layer $B$ reads
\begin{widetext}
\be
\frac{d{\mathbf{P}}_A}{dt}=-\frac{1}{V^3}
\!\!\!\sum_{{\mathbf{p}}, {\mathbf{p}}', {\mathbf{q}}}\!\!\!{\mathbf{p}}w({\bf q})\left[n_{\pp A} n_{\pp' B}
(1-n_{\mathbf{p-q} A})(1-n_{\mathbf{p'+q} B})-(\pp\leftrightarrow{\mathbf{p-q}},\;\pp'\leftrightarrow{\mathbf{p'-q}})\right]\delta(\epsilon_{\pp B}+\epsilon_{\pp' A}-\epsilon_{\mathbf{p-q} A}-\epsilon_{\mathbf{p'+q} B}),
\label{dPdt}
\ee
\end{widetext}
with $w({\mathbf q})$ the probability transition for the scattering process $\pp_A+\pp'_B\rightarrow (\mathbf{p-q})_A+(\mathbf{p'+q})_B$.
Taking the continuum limit and introducing the density-density response function for the layer $X=A,\;B$
\be
\chi_X(q,\omega)=\int\frac{d^2{\bf p}}{(2\pi)^2}\frac{n_{\pp,
X}-n_{{\mathbf{p+q}}, X}}{\epsilon_\pp-\epsilon_{\mathbf{p+q}}+\hbar\omega+i0^+}
\ee
Eq. (\ref{dPdt}) one gets
\begin{eqnarray}
\frac{d{\mathbf{P}}_A}{dt}&=&-\int\frac{d^2{\mathbf q}}{(2\pi)^2}{\mathbf{q}}\;w(\qq)\nonumber\\
&\times&\!\!\!\int_{-\infty}^{\infty}
d\omega\frac{
\rm{Im}\chi_A(q,\qq\cdot\vv-\omega)\rm{Im}\chi_B(q,\omega)}
{(1-{\rm e}^{\beta(\omega-\qq\cdot\vv)}(1-{\rm e}^{-\beta\omega})},
\label{dPdt3}
\end{eqnarray}
Since we are interested in the low temperature behaviour we can use the zero temperature response function in the previous expression. Moreover for the aim of the present paper we just use the golden rule for the collision probability with the bare dipolar potential, i.e., $w({\bf q})=2\pi |V(\qq)|^2/\hbar$ (but see text below).

Analytical expressions for the relaxation rate are possible in some limiting cases.
We are interested in particular in the dependence of $\tau_P$ on the various parameters at low-$T$ and low-$v$. 

\paragraph{Velocity dependence at $T=0$.}
Assuming zero temperature and low velocity such that $m v\ll k_{F,A}$, with $k_{F,A}$ the Fermi momentum in the layer $A$, we can use the expression ${\rm Im}\chi_{A(B)}\simeq m^2\omega/(2\hbar^3 qk_{F,A(B)})$\cite{JauhoSmith} in Eq. (\ref{dPdt3}). We can then write the momentum relaxation time as
\be
\frac{1}{\tau_P}=\frac{5}{24 \pi^2}\frac{mv^2}{\hbar}\left(\frac{r^*}{\lambda}\right)^2
\frac{1}{(\lambda k_{F,B})^4}\left(\frac{k_{F,B}}{k_{F,A}}\right)^3.\label{t0}
\ee

\paragraph{Temperature dependence at very low-$v$.}
In this case we find the leading term at small temperature. Such a result is usually the one of main interest in solid state physics. Assuming that the temperature is low enough that the gases in both layers are degenerate, but $vk_{F,A}<k_BT$, we can again use the low-frequency expression for ${\rm Im}\chi_{A(B)}$ and we obtain
\be
\frac{1}{\tau_P}=\frac{\pi}{32}\frac{(k_B T)^2}{\hbar^2\epsilon_{F,B}}\left(\frac{r^*}{\lambda}\right)^2\frac{1}{(\lambda k_{F,B})^2}\left(\frac{k_{F,B}}{k_{F,A}}\right)^3.\label{tT}
\ee

General arguments on the phase space restriction for the collisions suggest that the leading terms are indeed $v^2$ and $T^2$. Within our approximations we find a $1/\lambda^4$ dependence at finite temperature and at zero-temperature a $1/\lambda^6$ dependence.

It is worth noticing that the same finite temperature dependence on the layers' distance has been found for large enough inter-layer separation for a double layer electron gas \cite{JauhoSmith} and graphene \cite{SarmaGraphene} due to the screening of the Coulomb potential. Indeed the screened Coulomb potential within random phase approximation has the same functional form of the dipole inter-layer potential \cite{RMP2DEG}. In the random phase approximation the effective dipolar potential due to (static) screening in the layer-$B$ for exchange momentum $q<2 k_F$ reads
\be
V_{\rm RPA}({\bf q})=-\frac{2\pi d^2 q e^{-\lambda q} }{1-r^*q},
\ee
where we used the static value of the polarizability of a two dimensional ideal Fermi gas
Thus in our case (due to the different scaling of the dipolar potential with respect to the Coulomb one)
for large inter-layer separation, $r^*/\lambda\le 1$, we can consider the unscreened potential and find Eqs. (\ref{t0}) and (\ref{tT}).

\section{Considerations on experiments with ultra-cold trapped gases}

In the previous sections we consider an uniform two-dimensional gas. In most of the experiment on cold gases an external parabolic confinement is present. In particular a two dimensional configuration is realized provided the confinement is strong enough to freeze the motion in one direction, e.g., along $z$. For a cylindrical symmetric trapping the potential in the other remaining directions $x$, $y$ can be written  as
\be
V_{\rm ho}(x,y)=\frac{1}{2}m\omega_\perp (x^2+y^2).
\ee
Within Hartree-Fock approximation for the energy of the system the density profile of the gas in such a potential can be determined by a variational Thomas-Fermi ansatz which in 2D is simply an inverted parabola
\be
n(x,y)=\frac{2N}{\pi R_\perp^2}\left(1-\frac{x^2+y^2}{R_\perp^2}\right),
\label{LDA}
\ee
where $N$ is the number of atoms/molecules and $R_\perp$ is the variational radius.
We notice that since we consider that the dipoles are oriented perpendicularly to the plane we do not need to consider deformations of the Fermi surface (see, e.g., for the 2D case \cite{BruunTaylor,CDWSogo}).

A double layer configuration can also be realized, e.g., by means of selecting two wells in an one-dimensional optical lattice. For the sake of simplicity in the dscussion we assume that the two dimensional confining potential is the same for both layers.
 
The polaron energy depends on the density of the layer $B$ through $k_{F,B}$. Thus in a local density approach the polaron feels an effective external potential \be V_p(x,y)=V_{\rm ho}(x,y)+\epsilon_p^0(n_B(x,y))\ee
where we assume the density in the layer $B$, $n_B$ has the shape (\ref{LDA}).

If it is possible to excite the dipole mode for the polaron \cite{RecatiLobo}, i.e., the frequency of small oscillations of a dilute polaron gas, one will find a value of its frequency $\omega_p$ larger than the bare value $\omega_\perp$. In particular using the simple expression Eq. (\ref{E_int_bound}) one finds 
\be
\omega_p=\omega_\perp\left[1+\frac{|\epsilon_p^0|}{\epsilon_{F,0}}\left(1-\frac{1}{2}\frac{|\epsilon_p^0|}{\epsilon_{F,0}}\frac{1}{1+|\epsilon_b|/(2\epsilon_{F,0})}\right)\right]^{1/2},\label{dipfreq}
\ee 
where $\epsilon_{F,0}$ is the value of the Fermi energy at the center of layer $B$.
For small inter-layer interaction we have simply $\omega_p/\omega_\perp\simeq 1+\epsilon_p^0/\epsilon_{F,0}$ while it saturates for strong interaction.

In order to measure the frequency Eq. (\ref{dipfreq}) the dipole mode should decay slowly such that enough oscillations occur.
The damping time of such oscillation is related to the momentum relaxation time studied in Sect. V (as explained in \cite{GeorgAlessio}).
In particular if $\omega_p\tau_p\gg 1$ the dipole mode is collisionless and thus weakly damped, while if $\omega_p\tau_p\ll 1$ the mode is hydrodynamic and therefore overdamped.

It is worth mentioning that for a short-range interacting spin-1/2 Fermi gas the frequency of the quadrupole \cite{ENSquadrupole} and dipole \cite{Martin} polaron modes and their decay rates has been recently measured providing informations on the Landau parameters of the unbalanced normal phase of the gas \cite{RecatiLobo} and a new insight in the transport properties of strongly interacting Fermi system \cite{Martin}.

\section{Conclusion}

In conlusion, we calculated the energy of a dipolar polaron in a double layer configuration for relatively weak interactions assuming the bath being in its Fermi liquid phase. 
In the same spirit we determined the momentum relaxation rate of a (gas of) polaron as a function of the gas' speed and of the temperature of the system. The latter one is related to the well known drag effect in a double layer electronic structure. Furthermore, we estimated the frequency shift and decaying time of the out-of-phase or dipolar mode in a trapped configuration, which should correspond to a realistic experimental implementation of our system.

\acknowledgments
Useful discussions with Chiara Menotti, Alexander Pikovski and Luis Santos are acknowledged. 
This work has been supported by ERC through the QGBE grant.


\begin{thebibliography}{99}

\bibitem{Baranov2008}
M. A. Baranov, Physics Reports {\bf 464}, 71 (2008).

\bibitem{Lahaye2009}
T. Lahaye \emph{et al.}, Rep. Prog. Phys. {\bf 72}, 126401 (2009).

\bibitem{Ni2008} K.-K. Ni \emph{et al.}, Science {\bf 322}, 231 (2008).

\bibitem{Deiglmayr2008} J. Deiglmayr \emph{et al.}, Phys. Rev. Lett. {\bf 101}, 133004 (2008).

\bibitem{Ospelkaus2010} S. Ospelkaus {\it et al.}, 
Science {\bf 327}, 853 (2010).

\bibitem{Ni2010} K.-K. Ni {\it et al.}, 
Nature {\bf 464}, 1324 (2010).

\bibitem{Michael2D}
M. Klawunn, A. Pikovski and L. Santos,
Phys. Rev. A {\bf 82}, 044701 (2010).

\bibitem{Armstrong2010}
J. R. Armstrong et al., Europhys. Lett. 91, 16001 (2010).

\bibitem{Baranov2011}
M. A. Baranov, A. Micheli, S. Ronen, and P. Zoller,
Phys. Rev. A {\bf 83}, 043602 (2011).

\bibitem{BCS-bilayer}
A. Pikovski, M. Klawunn, G.V. Shlyapnikov, and L. Santos,
Phys. Rev. Lett. {\bf 105}, 215302 (2010).

\bibitem{Potter2010}
A. C. Potter, E. Berg, D.-W.Wang, B. I. Halperin, and E. Demler,
Phys. Rev. Lett. {\bf 105}, 220406 (2010). 

\bibitem{Pikovski2011}
A. Pikovski, M. Klawunn, A. Recati, and L. Santos,
eprint: arXiv: 1108.5642v1.

\bibitem{ExpPol}
A. Schirotzek, C.-H. Wu, A. Sommer and M. W. Zwierlein,
Phys. Rev. Lett. {\bf 102}, 230402 (2009);
S. Nascimbene, N. Navon, K. J. Jiang, L. Tarruell, M. Teichmann, J. McKeever, F. Chevy, and C. Salomon, Phys. Rev. Lett. {\bf 103}, 170402 (2009);
N. Navon, S. Nascimbene, F. Chevy, C. Salomon, Science {\bf 328}, 5979 (2010);
B. Fr\"ohlich, M. Feld, E. Vogt, M. Koschorreck, W. Zwerger, and M. K\"ohl, 
Phys. Rev. Lett. {\bf 106}, 105301 (2011).

\bibitem{ChevyMora}
F. Chevy and C. Mora, Rep. on Prog. in Phys. {\bf 73},   112401 (2010).

\bibitem{McGuire}
J. B. McGuire, J. Math. Phys. (N.Y.) 7, 123 (1966).

\bibitem{Roland1D}
S. Giraud and R. Combescot, Phys. Rev. A {\bf 79}, 043615 (2009).

\bibitem{var2D-1}
S. Z\"ollner, G. Bruun and C. J. Pethick,
Phys. Rev. A {\bf 83}, 021603 (2011).

\bibitem{var2D-2}
M. Parish, Phys. Rev. A {\bf 83}, 051603 (2011).

\bibitem{OurPRA}
M. Klawunn and A. Recati,
Phys. Rev. A {\bf 84}, 033607  (2011).

\bibitem{Schmidt2D}
R. Schmidt, T. Enss, V. Pietil\"a and E. Demler, 
eprint: arXiv: 1110.1649.

\bibitem{Alessio2007}
R. Combescot, A. Recati, C. Lobo and F. Chevy, Phys. Rev. Lett. {\bf 98}, 180402 (2007).
 

\bibitem{PilatiGiorgini}
S. Pilati and S. Giorgini, Phys. Rev. Lett. {\bf 100}, 030401 (2008).

\bibitem{Schmidt2011}
R. Schmidt and T. Enss,
Phys. Rev. A {\bf 83}, 063620 (2011).

\bibitem{Lipparini}
E. Lipparini,
{\em Modern many particle physics (2nd edition)},
World Scientific 2008.

\bibitem{inlayer}
C.-K- Chan, C. Wu, W.-C. Lee, and S. Das Sarma,
Phys. Rev. A {\bf 81}, 023602 (2010).

\bibitem{finitewidth}
For our considerations the influence of the finite width of the layers $l_z\ll\lambda$ is negligible.
However, in similiar situations it can lead to very interesting physics as discussed in the two recent preprints: M. Babadi, and E. Demler, eprint: arXiv: 1109.3755v2 and
V. Pietil\"a, D. Pekker, Y. Nishida, and E. Demler,
eprint: arXiv: 1110.0494v1.


\bibitem{BG1957}
H. A. Bethe and J. Goldstone,
Proc. Roy. Soc. A {\bf 238}, 551 (1957). 

\bibitem{Nagano1984}
S. Nagano, K. S. Singwi and S. Ohnishi,
Phys. Rev. B {\bf 29}, 1209 (1984).

\bibitem{Astra2007}
G. E. Astrakharchik, J. Boronat, I. L. Kurbakov and Yu. E. Lozovik,
Phys. Rev. Lett. {\bf 98}, 060405 (2007).

\bibitem{Giorgini2D}
Stefano Giorgini, private communications.

\bibitem{Rojo}
A. G. Rojo, J. Phys.:Condens. Matter {\bf 11}, R31 (1999).

\bibitem{SarmaGraphene}
W.-K. Tse, B. Y.-K. Hu and S. Das Sarma, Phys. Rev. B {\bf 76}, 081401(R) (2007) 

\bibitem{Polini}
M. Polini and G. Vignale, Phys. Rev. Lett. {\bf 98}, 266403 (2007);
R. A. Duine, M. Polini, H. T. C. Stoof, and G. Vignale,
Phys. Rev. Lett. {\bf 104}, 220403 (2010).

\bibitem{GeorgAlessio}
G. M. Bruun, A. Recati, C. J. Pethick, H. Smith and S.~Stringari, Phys. Rev. Lett. {\bf 100}, 240406 (2008).

\bibitem{JauhoSmith}
A.-P. Jauho and H. Smith, Phys. Rev. B {\bf 47}, 4420 (1993).

\bibitem{RMP2DEG}
T. Ando, A. B. Fowler and F. Stern, Rev. Mod. Phys. 54, 437 (1982).

\bibitem{BruunTaylor}
G. M. Bruun and E. Taylor, Phys. Rev. Lett. {\bf 101}, 245301 (2008).

\bibitem{CDWSogo}
Y. Yamaguchi, T. Sogo, T. Ito and T. Miyakawa, Phys. Rev. A {\bf 82}, 013643 (2010).

\bibitem{RecatiLobo}
 A. Recati, C. Lobo, and S. Stringari, Phys. Rev. A {\bf 78}, 023633
(2008).

\bibitem{ENSquadrupole}
S. Nascimb\`ene, N. Navon, K. J. Jiang, L. Tarruell, M. Teichmann, J. McKeever, F. Chevy, and C. Salomon, Phys. Rev. Lett. {\bf 103}, 170402 (2009). 

\bibitem{Martin}
A. Sommer, M. Ku, G. Roati, M. W. Zwierlein, Nature {\bf 472}, 201 (2011); A. Sommer, M. Ku, M. W. Zwierlein, New J. Phys. {\bf 13}, 055009 (2011).

\end{thebibliography}
\end{document}